# The Impact of Operators' Performance in the Reliability of Cyber-Physical Power Distribution Systems

Michel Bessani[1], Rodrigo Z. Fanucchi[2], Alexandre C.C. Delbem[3], Carlos D. Maciel[1*]

[1] Department of Electrical and Computer Engineering, University of São Paulo, São Carlos, Brazil
[2] Department of North Maintenance, Copel Distribution S/A, Londrina, Brazil
[3] Institute of Mathematical and Computer Sciences, University of São Paulo, São Carlos, Brazil
[*] carlos.maciel@usp.br

**Abstract:** Cyber-Physical Systems are the result of integrating information and communication technologies into physical systems. One particular case are Cyber-Physical Power Systems (CPPS), which use communication technologies to perform real-time monitoring and operation. These kinds of systems have become more complex, impacting on the systems' characteristics, such as their reliability. In addition, it is already known that in terms of the reliability of Cyber-Physical Power Distribution Systems (CPPDS), the failures of the communication network are just as relevant as the electrical network failures. However, some of the operators' performances, such as response time and decision quality, during CPPDS contingencies have not been investigated yet. In this paper, we introduce a model to the operator response time, present a Sequential Monte Carlo Simulation methodology that incorporates the response time in CPPDS reliability indices estimation, and evaluate the impact of the operator response time in reliability indices. Our method is tested on a CPPDS using different values for the average response time of operators. The results show that the response time of the operators affects the reliability indices that are related to the durations of the failure, indicating that a fast decision directly contributes to the system performance. We conclude that the improvement of CPPDS reliability is not only dependent on the electric and communication components, but also dependent on operators' performance.

## 1. Introduction

Nowadays, the majority of engineered systems have been integrated with Information and Communication Technologies (ICTs) [1, 2], e.g., Marine Ships [3]. Additionally, modern society is increasingly reliant on engineered systems that are critical infrastructures, these include Water Distribution Systems [4] and Energy Systems [5], which should provide products and services in a reliable way [6].

The integration of ICTs into physical and engineered systems defines Cyber-Physical Systems (CPSs) [2]. CPSs are networked systems with actuators, sensors, and processors designed to interact with physical components, including the system user or operator [4]. These components are used to perform system operation using real-time information and control [7]. Despite these benefits, such systems are becoming even more complex and consequently, present new challenges concerning representation, modelling and quantification [1, 8].

Systems reliability is one of the CPSs challenges, as the number of components and connections in the system increases they also become more interdependent [1, 9]. This dependence between components makes the CPS more susceptible to cascading failures [10], i.e., a failure of one or more elements of the cyber system can cause a failure in others components of the physical system. Reliability of power systems becomes an area of major focus [11] once the Smart Grid (SG) philosophy [12, 13] is leading to even more complex Cyber-Physical Power Systems (CPPSs).

CPPSs raises the need for estimating the reliability of the two parts of the system operating simultaneously. A failure of the cyber network can cause erroneous measures of physical parameters and unrealized operator commands due to communication failure. Kirschen and Bouffard's paper [14] summarizes several blackouts that started with information and communication network failures. In particular, the Italian blackout of 2003 and the Western and Central Europe incident of 2006 did not involve electrical components failures.

During Power Distribution Systems contingencies, the system topology is dynamically changed to either solve or minimize contingency impacts, i.e., isolate the faulted parts and restore as many out-of-service healthy areas as possible [14]. Such service restoration is performed in distribution operation centers, and the decision-making process is based on the expertise and knowledge of each human operator [15]. Consequently, the human operators performance directly affects the network operation [16].

Decision-Support Systems (DSSs) [17] have been proposed to improve operators performance during contingencies [14, 18]. The CPPS real-time monitoring and physical measures allow the development of Intelligent Systems that can infer during a contingency situation and help the distribution operator decision-making process. Current reliability analysis of CPPS contemplates the integration of ICT in the Power System [19–24]. It is already known that Cyber network failures directly reduces the reliability of the CPPS, e.g., increases the loss of load probability [19].

However, the impact of operator performance during service restoration has not yet been investigated from a reliability viewpoint. During contingencies, the decision on service restoration is left to the operators, who already have other obligations [16]. Such decision making varies in quality and in time to be taken, and need to be embedded in modelling and simulation of CPS [25].

In this paper, we present a Sequential Monte Carlo Simulation methodology to estimate Cyber-Physical Power Distribution Systems (CPPDS) reliability indices incorporating the operators' response time and considering real-time network monitoring and operation. The operator response time is embedded in CPPDS failures simulation. The main contributions of this paper are:

- Propose a stochastic model that reflects the human operators response time;
- Introduce a simulation framework for reliability analysis of CPPDS that incorporates the operator's response time during service restoration;
- Evaluate the impact of such operator response time on the CPPDS reliability.

The remainder of this paper is organized as follows: Section 2 presents information about CPPS and CPPDS service restoration, Section 3 shows the fundamentals of reliability analysis for CPPS and defines the human operator model, in Section 4 we introduce the proposed methodology, in Section 5 we perform a case study of our methodology, and Section 6 concludes this paper.

## 2. Cyber-Physical Power Systems

The Power Grid (Generation, Transmission and Distribution) are probably one of the most complex systems ever made [26]. In the USA [27], the electricity system has hundreds of generating units, hundreds of transmission lines, millions of transformers and hundreds of millions of protection and control elements. The transmission system energizes distribution feeders which delivers energy to the customers at buses through transformers [28].

Some characteristics of the conventional power grid are one-way communication, few sensors, manual monitoring and restoration [29]. In contrast, the Smart Grid [13] is a power grid with an ICT network, allowing two-way communications, many sensors, self-monitoring, adaptive and islanding topologies. Such differences between the conventional and the SG aim to implement some properties into the power system [30], e.g., self-healing, high reliability and power quality, resistance to cyber-attacks and optimized monitoring and operation.

ICT can be divided into some main categories [22]:

- Data Acquisition: real-time data acquisition from important parts of the power grid, such as voltage at the buses.
- Communication: responsible for delivering the acquired data to the information processing and for transmitting control signals to the actuators on the power grid.
- Information Processing: processes the data to provide system information to the operator, it consists of monitoring tools and analysis software that provide reliable information about the power grid.
- Graphical User Interface: displays all the power grid information collected and processed by the operator.

Any failure of these parts is crucial for the operator decision-making during system contingency. It can cause incorrect system operation, e.g., a control signal can be not performed by and actuator. The ICT can provide erroneous measures or information about the current system state, or the ICT network may suffer a cyber-attack. Any of the mentioned situations can lead to a system failure, or a blackout, as exemplified in [31].

*2.1. CPPDS service restoration*

During CPPDSs contingencies [14] the faulted components are identified by the system operators through sensors between the system branches, the faulted branch are isolated by opening sectionalizing switches (normally-closed), and then, the operator reenergizes the healthy out-of-service branches after the disconnected one by closing tie switches (normally-open). The performance of isolation and restoration during contingencies is dependent of the switches that divide the feeders into branches, and the electrical capability of the system to restore the healthy branches after the faulted one [32].

The service restoration at distribution grid can be treated as a network reconfiguration [15], where the determination of a solution is a combinatorial problem [33]. Furthermore, the service restoration can be focused on different objectives [14, 15, 33], e.g., minimize the number of switching operations and out-of-service customers. Additionally, changes in the system topology need to respect the system radiality and electrical constraints [14]. In summary, the service restoration is a multi-objective and multi-constrained problem that is solved by human operators at the distribution operation centers.

The importance of human operators performance in power system have already been highlighted [16, 25, 34]. Any human has limited cognitive functions, e.g., reasoning and memory capacities [16, 35], such limitations can affect the time to define the solutions as well as the solutions quality.

## 3. Reliability Analysis of Cyber-Physical Power Systems

Reliability Analysis aims to investigate systems operations and failures [36]; it is performed considering system configuration, redundancies, and the reliability data of system components. The reliability analysis is becoming more important since the electrical market is becoming even more competitive, and customers are more demanding with the quality of energy delivered.

For evaluation of power system reliability, two main groups of techniques are used [20]: analytical methods, and Monte Carlo Simulation. The first one uses mathematical models and numerical algorithms to evaluate the system reliability and obtain the reliability indices, e.g., Falahati and Fu paper [19]. On the other hand, Monte Carlo Simulation (MCS) is a virtual experiment that uses the stochastic aspect of system components to run a simulation of system behavior over a given time.

MCS can be divided into two approaches: Pseudo-Sequential Monte Carlo Simulation [37] and Sequential Monte Carlo Simulation (SMCS) [36]. Pseudo-Sequential Monte Carlo Simulation [37] samples some states of the system using non-sequential simulation and simulate some neighboring states, these samples are used to obtain the reliability indices estimation. The SMCS generates a synthetic history of system failures and repairs respecting the failure rates and repair times of each system's component. The advantage of SMCS over Pseudo-SMCS is that we can repeat the simulation to obtain the estimation and the probability distribution of the reliability indices [38].

### 3.1. Sampling Failure and Repair Times

We can consider that CPPDS branches and communication network components are a two-state system (working and failure) in which transition probabilities from one state to the other depend only on the present state, also known as a system with "lack-of-memory" [36]. Assuming that the failure rate (1) of a CPPDS branch is constant, i.e., after a failure, it can be considered new due to maintenance or replacement of the failed component, it can be modeled using an Exponential Reliability function (2) as follows:

$$\lambda = \frac{number\ of\ failures}{total\ operation\ time\ of\ units}, \tag{1}$$

$$R(t) = \epsilon^{-\lambda t}, \tag{2}$$

where $\lambda$ is the failure rate, and $R(t)$ is the Reliability function. The Reliability function can be defined as [39] the probability of a system functioning properly during a specified time interval with the start from time zero (3).

$$R(t) = P(T > t) \tag{3}$$

We can obtain the cumulative distribution function (4), and the probability distribution function (5) using (2) and (3) as shown below:

$$P(t) = P(T \leq t) = 1 - P(T > t) = 1 - R(t), \tag{4}$$

$$f(t) = \frac{\partial P(t)}{\partial t} = \lambda e^{-\lambda t}. \tag{5}$$

Knowing that the cumulative distribution function (c.d.f) assumes values between 0 and 1 for different $t$ values, we use the Inverse Transform Method [28, 36]:

$$U = P(t) = 1 - \epsilon^{-\lambda t}, \tag{6}$$

where $U$ is the uniform distribution on the interval [0,1]. We can rewrite (6) as (7):

$$t = -\frac{1}{\lambda}\ln(1 - U), \tag{7}$$

where $(1 - U)$ also is the uniform distribution on the interval [0,1], and $t$ follows the exponential distribution (2). Finally, (7) can be rewritten as (8):

$$t = -\frac{1}{\lambda}\ln(U) \tag{8}$$

Using (8) we can perform random sampling to simulate the system branches' times to fail. Considering that the repair time ($t_r$) follows a Normal distribution, $t_r \sim N(\mu_r, \sigma_r^2)$, we use the Box-Muller method [40] to sample $t_r$ values for each branch failure, as follows:

$$X = \sqrt{-2 \ln U_1} \cos(2\pi U_2), \tag{9}$$

where $U_1$ and $U_2$ are uniform distributions in the interval [0,1] and $X \sim N(0,1)$, to obtain $t_r \sim N(\mu_r, \sigma_r^2)$ we perform as (10):

$$t_r = X\sigma_r + \mu_r, \tag{10}$$

*3.2. Human Operator model*

Traditionally, SMCS consider the failures of electrical and communication components [19–24]. In order to simulate the human operator performance, we introduce a random variable to represent the time that takes an operator to define a solution during contingencies and we call it the *Response Time of Operator* (*RTO*). The random variable was chosen to enable the sampling of RTO during the SMCS, similar to time to failure and time to repair. A Normal probability distribution was adopted to describe the RTO:

$$t_{RTO} \sim N(\mu_{RTO}, \sigma_{RTO}^2) \tag{11}$$

The Normal distribution mean ($\mu_{RTO}$) represents the average *RTO*, and its deviance ($\sigma_{RTO}^2$) represents both the best and the worst response times. The Box-Muller method [40] were used to sample the RTO for each failure, similar to repair time:

$$X = \sqrt{-2 \ln U_1} \cos(2\pi U_2), \tag{12}$$

$$t_{RTO} = X\sigma_{RTO} + \mu_{RTO} \tag{13}$$

*3.3. Obtaining a State Transition Vector*

A state transition vector [28] describes the transitions between the working and failure states. It is obtained by using the sampling methodologies presented in Subsections 3.1 and 3.2. It is a synthetic history of system failures and repairs respecting the failure rates and repair times of each system part, and is obtained using the following steps:

1. Define a duration length of the state transition vector ($T$) and start the transition vector at time 0 with the part of interest functioning properly;
2. Sample the time to failure using equation (8) and add it to the vector;
3. Sample the repair time using equations (9) and (10) and add the repair duration to the vector;
4. Repeat steps 2 and 3 until the time $T$ is reached.

## 4. Proposed Methodology

Our methodology consists of an SMCS that simulates the stochastic behavior of the following CPPDS parts: electrical components failures, communication network components failures and the system operation's response time during contingencies. We consider as system failure the states that have customers without energy.

Figure 1 illustrates our SMCS. The blue blocks represent the simulation of the cyber network failures. A state transition vector is computed for each communication component. If a communication switch fails, we will try to perform an optical switches path reconfiguration applying the Rapid Spanning Tree Protocol (RSTP) [41], which takes thirty seconds to reconfigure the communication paths, this time is incorporated in the state transition vector. When a controller fails the SMCS will sample a repair time.

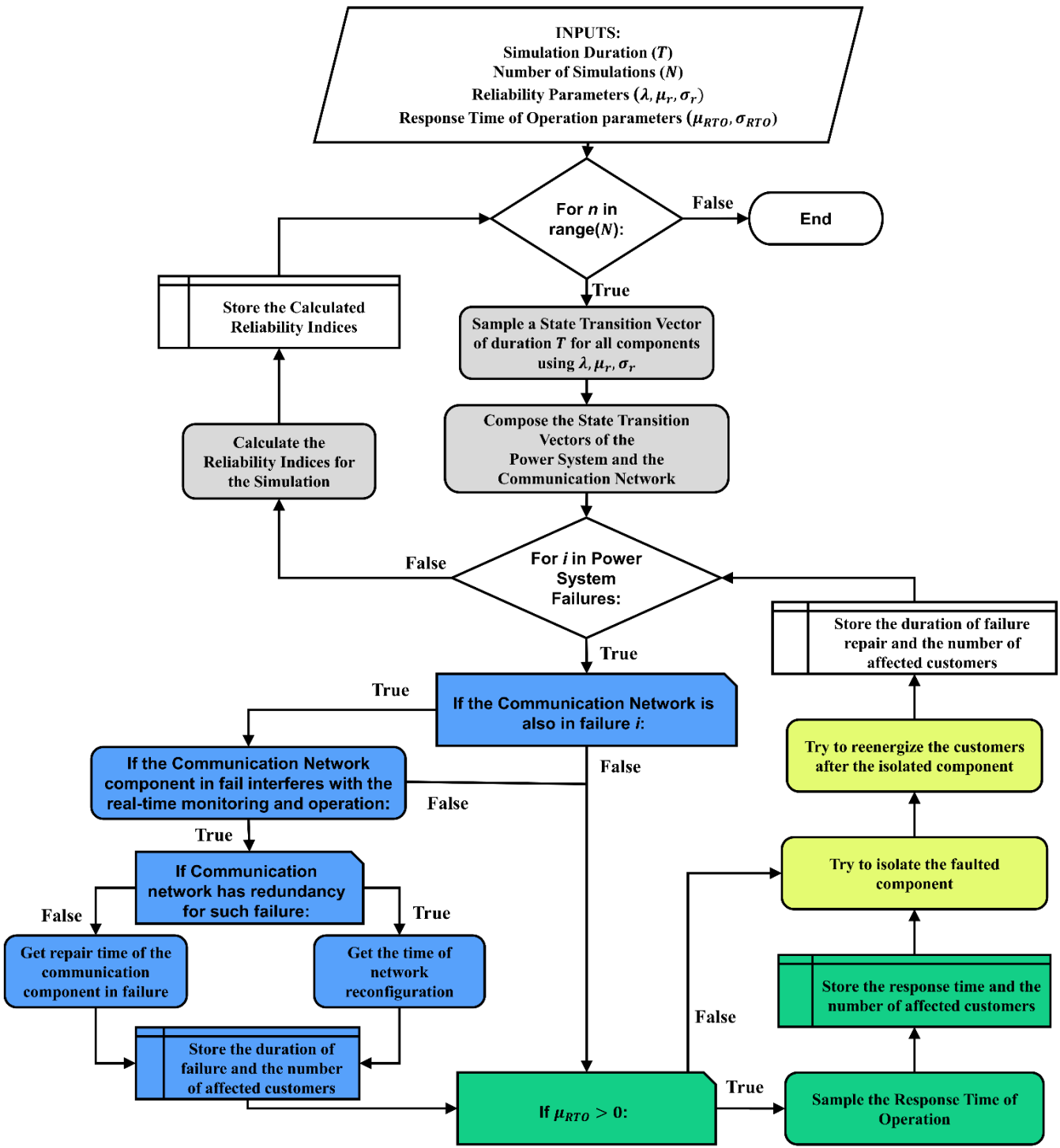

Figure 1 – Proposed Sequential Monte Carlo Simulation methodology to incorporate electrical and communication components failures and the response time of operator in reliability indices of Cyber-Physical Power Distribution Systems. The yellow symbols represent the changes in the system topology for power restoration in healthy branches during contingencies. The blue symbols represent the impact of communication networks failures. The green symbols compute the operation response time impacts.

The yellow blocks represent the process of isolate faulted branches and reenergized the healthy out-of-service branches by opening sectionalizing switches and closing tie switches, this is performed considering the electrical connectivity computed using a graph representation [33]. The green blocks compute the impact of the operator performance (RTO), if $\mu_{RTO} > 0$ we sample $t_{RTO}$ and consider that all the customers at the branches affected during contingency stay without energy during $t_{RTO}$.

We obtain samples of the reliability indices using the duration and quantity of customers affected by each failure during one simulation. Repeating the simulation $N$ times, we obtain a set of samples that provide an estimation of the real value of the CPPS reliability indices. The SMCS calculates the following reliability indices [42]: Failure Rate, Availability, SAIDI, and SAIFI, defined in Table 1.

Table 1 - Reliability indices calculated by SMCS [42]. **Failure Rate and SAIFI are related to the failures occurrences, and Availability and SAIDI are related to the duration of the failures.**

| Reliability indices | Equation | Description |
| --- | --- | --- |
| Failure Rate | $\lambda = \dfrac{Quantity\ of\ failures}{Time\ simulated\ (year)}$ | Average probability of system failure per year (failures/year) |
| Availability | $A = \dfrac{Time\ functioning}{Time\ simulated\ (year)}$ | Average percentage of time the system is functioning per year |
| SAIDI | $SAIDI = \dfrac{\sum_i U_i N_i}{\sum_i N_i}$ | System Average Interruption Duration Index (hours/system customer/year) |
| SAIFI | $SAIFI = \dfrac{\sum_i \lambda_i N_i}{\sum_i N_i}$ | System Average Interruption Frequency Index (interruption/system customer/year) |

The Failure Rate represents the probability of failure for a system during a year, and it is only related to the occurrence of failure states. Whereas SAIDI and SAIFI are related to the impact of faults on system customers and indicate respectively, average duration and average frequency of failures weighted by the number of clients affected. Availability represents the percentage of time that the system works correctly during a year, and in a system with a high reliability [12], it can be measured by the "number of nines" (14).

$$N_9 = -\log_{10}(1 - A) \qquad (14)$$

## 5. Methodology Verification - Case Study

We applied our SMCS to the Civanlar's distribution network [43], a monophasic three-feeder example distribution system obtained in the Repository of Distribution Systems (REDS) [44]. It consists of 13 branches, 3 feeders, 13 sectionalizing switches and 3 tie-switches, as shown in Figure 2. It also presents fifteen feasible possibilities of topological reconfiguration (open a sectionalizing switch and closing a tie switch) respecting the radiality and without isolating any branch, the total number of possibilities of switching options is much larger than fifteen.

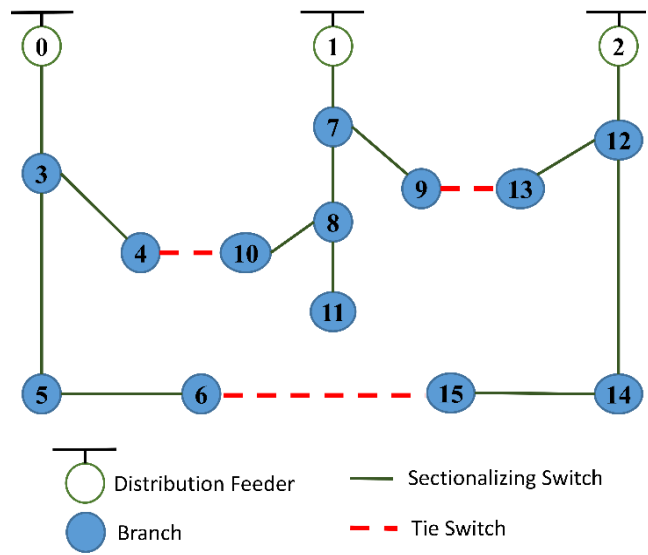

**Figure 2 – Electrical topology of the Civanlar's distribution network [43], which are a monophasic three-feeder example system. Its data is available in the Repository of Distribution Systems (REDS) [44]. It consists of 13 branches, 3 feeders, 13 sectionalizing switches and 3 tie-switches.**

An ICT network was added to the system respecting the Falahati [24] conclusions about fiber-optic communication network topology impacts into CPPS reliability. The ICT system is connected in a 1-ring topology and consists of 13 communication switches, 13 controllers, and two servers. We assumed that all the sectionalizing and tie switches are automated (time to reconfiguration equals to zero), and the controllers are responsible for monitoring branches and changing the states of the electrical switches. The system topology is presented in Figure 3.

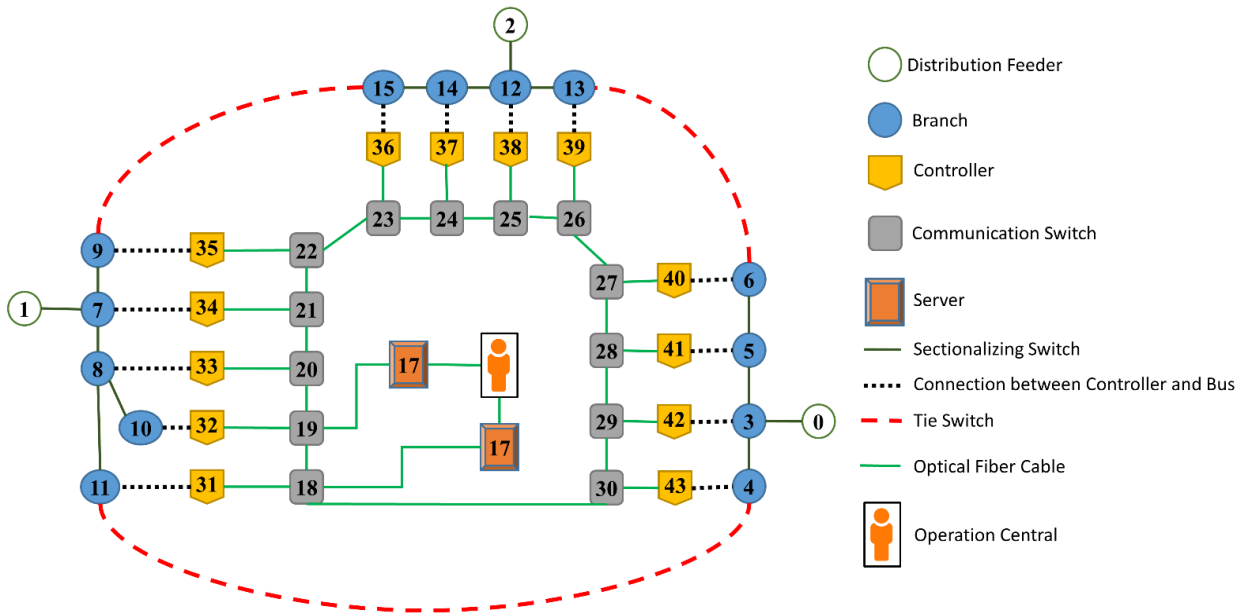

**Figure 3** – Representation of the Cyber-Physical Power System used in this paper. The Communication network has a 1-ring topology, each communication switch is connected to two others, allowing network reconfiguration in case of failures using the RSTP protocol [41].

The assumed quantities of customers at each network branch are shown in Table 2. The failure rates used for the CPPDS components are given in Table 3, and they were chosen considering the typical distribution equipment outage statistics presented by Chowdhury and Koval [39]. The distribution feeders, fiber-optic cables, servers, and switches are assumed to be fully reliable.

Our methodology presupposes that the distribution feeders are robust enough to handle the load transfer between each other during contingencies. All the cases were simulated using $T = 1000 \; years$ and $N = 1000$. We used the "number of nines" presented in the Equation (14), Section 4 to describe the results of Availability.

**Table 2** – Quantity of customers energized by each Cyber-Physical Power System branch.

| Branch id | Customers |
|---|---|
| 3 | 3 |
| 4 | 5 |
| 5 | 3 |
| 6 | 2 |
| 7 | 6 |
| 8 | 8 |
| 9 | 1 |
| 10 | 1 |
| 11 | 7 |
| 12 | 1 |
| 13 | 1 |
| 14 | 1 |
| 15 | 3 |

**Table 3 – Failure Rates ($\lambda$) and repair time parameters in hours ($\mu_r$ and $\sigma_r$) adopted to perform this study.**

| Component | $\lambda$ (failures/year) | $\mu_r$ (repair time - hours) | $\sigma_r$ (repair time - hours) |
|---|---|---|---|
| Branch | 0,100 | 3 | 0,6 |
| Communication Switch | 0,005 | 3 | 0,6 |
| Controller | 0,010 | 3 | 0,6 |

First we simulated two different cases to evaluate communication and electrical components failures impact in the reliability of CPPDS:

I. Power Grid with a fully reliable Cyber network and $\mu_{RTO} = 0$: The system has a fully reliable automated electrical switches and real-time monitoring, and the operator decision is instantaneous.

II. Power Grid with failures into Cyber network and $\mu_{RTO} = 0$: The same as the case above, but communication network can fail, resulting in a delay effect in system operation.

This first test of our SMCS methodology is to validate the capability of measuring the impact of Cyber network failures in the reliability indices. The results estimated are presented in Table 4.

**Table 4 – Reliability index using two different cases to evaluate the impact of communication and electrical components failures into the reliability of the studied Cyber-Physical Power System.**

| | Reliability Index | | | |
|---|---|---|---|---|
| Case | Availability (number of nines) | SAIDI (hours/system customer/year) | SAIFI (interruption/system customer/year) | Failure Rate (failure/year) |
| Power Grid with a fully reliable Communication network | 3,386 | 0,457 | 0,152 | 1,200 |

| Power Grid with failures into Communication network | 3,203 | 0,953 | 0,318 | 1,831 |

Table 4 shows that our methodology correctly incorporates the communications failures impacts to the CPPS reliability indices. The network reliability parameters are better in cases *I* than *II* due to the assumption of a communication network fully reliable. The observed decline in reliability reflects the importance of integrating a reliable communication network into the Power System.

A Cyber network failure can insert a delay during contingency situations, seeing that the distribution operation center will receive the information about contingency only if the communication network is functioning properly. Consequently, the duration of the average length of customers' power interruptions increases.

To analyze the impact of operation performance on the reliability of the CPPDS, we created different scenarios for the RTO and simulated them using the Power Grid with failures in the Communication network case. The RTO parameters used are presented in Table 5, each one represents different performances of a human operator making a decision during a system contingency.

Table 5 – Values of $\mu_{RTO}$ and $\sigma_{RTO}$ used to simulate the response time of operator.

| Response Time of Operator | |
|---|---|
| $\mu_{RTO}$ (minutes) | $\sigma_{RTO}$ |
| 0 | 0 |
| 1 | 0.2 |
| 5 | 1 |
| 10 | 2 |
| 20 | 4 |
| 40 | 8 |
| 60 | 12 |

The reliability indices estimated using our SMCS considering the parameters in Table 5 for RTO are presented in Table 6.

Table 6 - Reliability indices obtained for the CPPS with failures into Communication and Electrical components using the different average RTO values described in Table 5.

|  | Reliability Index | | | |
| --- | --- | --- | --- | --- |
| $\mu_{RTO}$ | Availability (number of nines) | SAIDI (hours/system customer/year) | SAIFI (interruption/system customer/year) | Failure Rate (failure/year) |
| 0 | 3,202699 | 0,953233 | 0,317813 | 1,831448 |
| 1 | 3,201778 | 0,956284 | 0,335595 | 1,899846 |
| 5 | 3,198109 | 0,968502 | 0,335595 | 1,899846 |
| 10 | 3,193573 | 0,983746 | 0,335595 | 1,899846 |
| 20 | 3,184608 | 1,014060 | 0,335595 | 1,899846 |
| 40 | 3,167266 | 1,075147 | 0,335595 | 1,899846 |
| 60 | 3,150581 | 1,136235 | 0,335595 | 1,899846 |

The increase of the response time directly affects the reliability indices that accounts failures durations – Availability and SAIDI, as the impact of a fast operation is related to the length of customers' power interruptions. When the $\mu_{RTO} > 0$, SAIFI and Failure Rate are affected because customers in the healthy out-of-service branches enter in failure state during the human operator's decision making. However, the impact is the same for any $\mu_{RTO} > 0$, since Failure Rate and SAIFI indices are related only to outages occurrence and are not influenced by the outages duration.

Table 7 presents the percentage of these impacts. The indices that consider only failure occurrence are not affected by the response time since the number of failures is not either. It is important to highlight that this study investigates the response time of the operator's impact on CPPS reliability; we do not analyze the effects of the operator's decision quality.

**Table 7 – Percentage difference in Availability and SAIDI indices when increasing the average RTO parameter on our methodology.**

|  | Reliability Index | | | |
| --- | --- | --- | --- | --- |
| $\mu_{RTO}$ (minutes) | Availability (number of nines) | %Availability | SAIDI (hours/system customer/year) | %SAIDI |
| 0 | 3,202699 | - | 0,953233 | - |
| 1 | 3,201778 | 0,028761 | 0,956284 | 0,320065 |
| 5 | 3,198109 | 0,114583 | 0,968502 | 1,277722 |
| 10 | 3,193573 | 0,141847 | 0,983746 | 1,573916 |
| 20 | 3,184608 | 0,564858 | 1,014060 | 6,381158 |
| 40 | 3,167266 | 1,106352 | 1,075147 | 12,789561 |
| 60 | 3,150581 | 1,627311 | 1,136235 | 19,198040 |

The impact of considering $\mu_{RTO} = 1$ is less than 1% for both SAIDI and Availability. On the other hand, the effect of considering $\mu_{RTO} > 40$ in Availability is more than 1%. The impact in SAIDI is greater than 1% for $\mu_{RTO} = 5$ minutes, and for $\mu_{RTO} = 60$, the impact is almost 20%. Such impacts are also shown in Figure 4. It indicates that a quick response time of operators is essential to the CPPDS reliability.

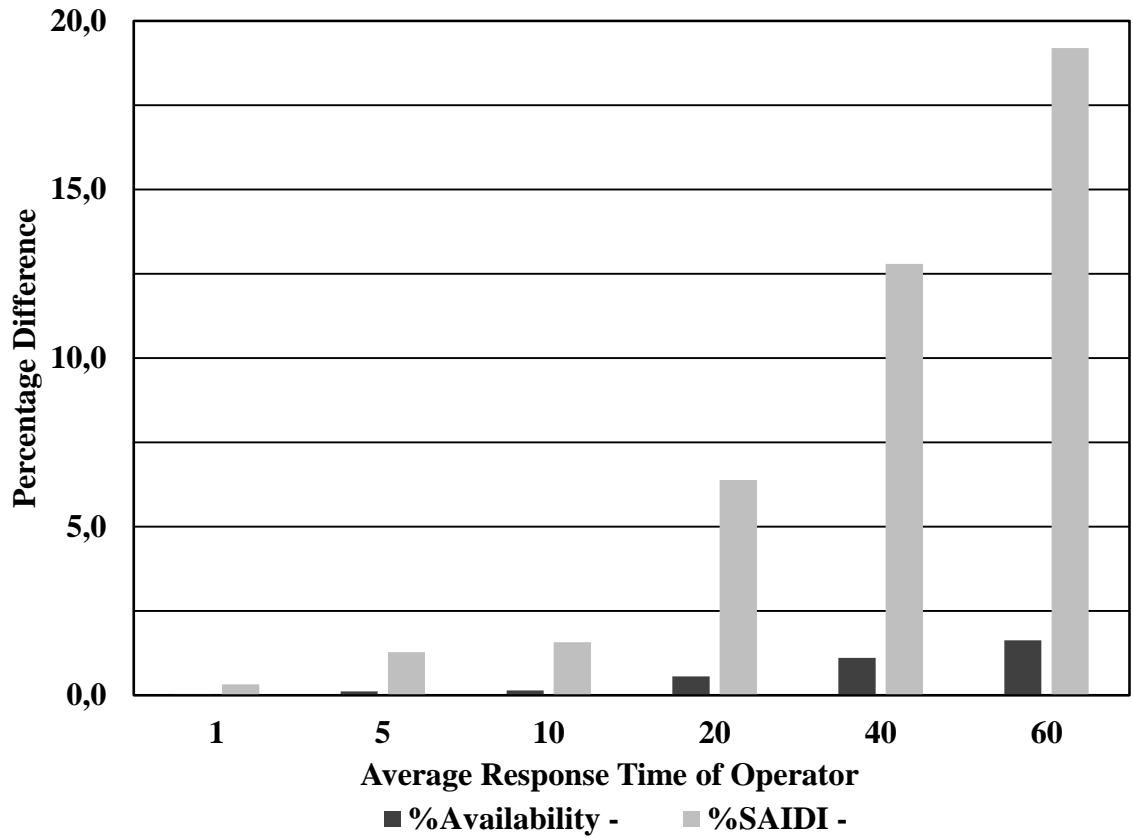

**Figure 4 – Percentage difference in CPPS Availability and SAIDI indices by the average RTO used in each scenario. The greater the average response time values, the larger the effect on reliability indices.**

## 6. Conclusions

In this paper, we presented a stochastic model to enable the incorporation of human operator's performance in a Sequential Monte Carlo Simulation that integrates electrical and communication components failures, and the response time of human operators in the calculus of Cyber-Physical Power System reliability indices. Different average response time values were used for simulation and the reliability indices obtained were analyzed.

Our results show the capability of our model to simulate the stochastic behavior of a Cyber-Physical Power System considering electrical and communication components failures, together with the response time of human operators. Moreover, the results indicate that the average response time of operator affects the reliability indices related to power interruptions durations. It also indicates that the average response time directly affects the reliability indices, showing that not considering such variable results in a bias of the reliability estimators.

In our case study, the RTO effect in Availability and SAIDI indices is linear, which is a consequence of the small size of our test system. The simulation generated failures with large time interval between them and without multiple failures. For a large-scale system, multiple failures will occur and the time interval between failures will be shorter, causing a chain effect for large response time, i.e., while the operators are making the decision for a failure another failure may happen.

The main advantage of our methodology is the capability to incorporate the human operator's performance considering the time necessary to make a decision. This time is present in every decision during contingencies since the human operator needs to consider many variables during their decision-making process.

In systems that do not have automated switches, the impact of the operators' response time will be reduced in comparison to our case study because the time to operate the switches is greater (we considered all the switches automated and with quick operation after the operator decision). The switches' operation time are affected by different factor, such as the number of crews to operate the manual switches and the area covered by the operation center.

Such findings reinforce the importance of Decision Support Systems development to the operation of Cyber-Physical Power Systems. CPPS are becoming even more reliable by using Information and Communication Technologies. Once the response time of the system operators also needs to be improved by such integration, investments in operators training and knowledge are necessary to improve performance and allow the adoption of new Decision Support Systems technologies.

It is important to mention that we only considered the time of decision-making. Our simulation assumes that the operation procedure always leads to the best configuration during contingencies. Incorporating the quality of the operator's decision may lead to additional findings on the impact of human operator's performance to Cyber-Physical Power Distribution Systems reliability. Application of the proposed methodology in real CPPDS can improve the capacity to estimate the reliability indices and help in directing investments to improve overall system reliability.

## 7. Acknowledgments

This paper is part of a project of Research and Development of Brazilian Electricity Regulatory Agency (PD 2866-0272/2012). The authors are also grateful to financial support from the National Council of Technological and Scientific Development (CNPq Brazil – Process Nº 475064/2013-5).